\documentclass[conference]{IEEEtran}
\IEEEoverridecommandlockouts

\usepackage{cite}
\usepackage{amsmath,amssymb,amsfonts}
\usepackage{algorithmic}
\usepackage{graphicx}
\usepackage{textcomp}
\usepackage{xcolor}
\usepackage{url}
\def\BibTeX{{\rm B\kern-.05em{\sc i\kern-.025em b}\kern-.08em
    T\kern-.1667em\lower.7ex\hbox{E}\kern-.125emX}}
\begin{document}

\title{A Unified Framework for Collecting Text-to-Speech Synthesis Datasets for 22 Indian Languages}

\author{\IEEEauthorblockN{Sujitha Sathiyamoorthy}
\IEEEauthorblockA{\textit{Dept of Computer Science \& Engineering} \\
\textit{Indian Institute of Technology Madras}\\
Chennai, India \\
sujivk16@gmail.com}
\and
\IEEEauthorblockN{N Mohana}
\IEEEauthorblockA{\textit{Dept of Computer Science \& Engineering} \\
\textit{Indian Institute of Technology Madras}\\
Chennai, India \\
n.mohanaravi@gmail.com}
\and
\IEEEauthorblockN{Anusha Prakash}
\IEEEauthorblockA{\textit{Independent Researcher} \\
Bengaluru, India \\
anushaprakash90@gmail.com}
\and
\IEEEauthorblockN{Hema A Murthy}
\IEEEauthorblockA{\textit{Dept of Computer Science \& Engineering} \\
\textit{Indian Institute of Technology Madras, Shiv Nadar University}\\
Chennai, India \\
hema@cse.iitm.ac.in}
}

\maketitle

\begin{abstract}
The performance of a text-to-speech (TTS) synthesis model depends on various factors, of which the quality of the training data is of utmost importance. Millions of data are collected around the globe for various languages, but resources for Indian languages are few. Although there are many efforts involved in data collection, a common set of protocols for data collection becomes necessary for building TTS systems in Indian languages primarily because of the need for a uniform development of TTS systems across languages. In this paper, we present our learnings on data collection efforts' for Indic languages over 15 years. These databases have been used in unit selection synthesis, hidden Markov model based, and end-to-end frameworks, and for generating prosodically rich TTS systems. The most significant feature of the data collected is that data purity enables building high-quality TTS systems with a comparatively small dataset compared to that of European/Chinese languages.   
\end{abstract}

\begin{IEEEkeywords}
Text-to-speech synthesis, data collection, Indian languages, unified protocol, IndicTTS23
\end{IEEEkeywords}

\section{Introduction}
\label{sec:intro}

The objective of this work is to develop high-quality datasets for training TTS synthesis systems in 22 Indian languages, focusing on a unified framework for data collection. The current text-to-speech systems  have evolved over a duration of more than 40 years. Popular frameworks for training TTS systems include USS, Hidden Markov Model (HMM) based and neural-network based approaches. With the advent of E2E architectures, training TTS systems has become easier using $<$text, audio$>$ pairs \cite{tacotron2, fastspeech2}. With the availability of many open-source toolkits for training, there has been widespread interest in academia and industry, leading to rapid advancements.

The achievement of a near-human synthesis requires high-quality TTS data consisting of audio recordings and corresponding transcriptions. The upper bound is largely dependent on the quality and accuracy of the dataset. Hence, data scraped from the web and other sources do not yield good results compared to using studio-recorded data with accurate transcriptions for training. Studies in \cite{Jeena_transcription_2019} show that having erroneous data in training degrades the synthesis output.

The performance of an automatic speech recognition (ASR) system is based on the word error rate (WER) on a test set. The WER increases proportionally to the mistakes made by the ASR system. On the other hand, TTS is a more challenging task, being a one-to-many problem. During subjective evaluations of TTS systems, it is often observed that even a small mistake in the synthesised audio leads to a disproportionately large drop in the evaluation score. In speech recognition, the visual system is the consumer, while in speech synthesis, it is the auditory system. The visual system has the ability to adapt to blur and can easily fill in the gaps in words or errors in words \cite{Thomas1986},  while the auditory system, which is sensitive to frequency and has a very high temporal resolution, cannot tolerate errors even in units as short as a mispronounced phoneme. Therefore, while ASR systems can tolerate a $4-6\%$ error in WER, TTS systems require $0\%$  WER.
Hence, it is essential that high-quality TTS systems are developed. TTS systems typically require studio-recorded data with accurate transcriptions. TTS data collection is expensive and time-consuming as this process  includes text collection, speech recording by a professional, and manual verification. Further, the speech should be delivered at a uniform rate.

According to Ethnologue\footnote{\url{https://www.ethnologue.com/}}, there are more than 7000 languages in the world. However, TTS systems have focused on very few languages, including English, Mandarin and European languages. There are more than 1369 rationalised languages in India \cite{censusIndia2011}, with 23 official languages, including English. We still do not have TTS systems for all the official Indian languages. Though there have been individual efforts for collecting data in a few languages \cite{srivastava-etal-2020-indicspeech}, a large-scale initiative for TTS data collection for Indian languages is limited \cite{ArunResources2016, LIMMITS_2023, rasa24_interspeech}.

The key contributions of our work are highlighted here:
\begin{itemize}
    \item We present a unified framework for data collection to develop Indian language TTS systems, specifically in terms of recording script collection, speaker selection, studio environment, speech recording and quality control.
    \item IndicTTS23 database: The aim is to collect 40 hours of single-speaker studio-recorded data (native male, female and corresponding English) in all 22 languages, amounting to 880 hours of data (765 hours collected so far).
    \item While good quality read speech TTS systems are available today, spontaneous speech synthesis is a requirement today.  Our early efforts on ``conversational-style TTS systems''  show that conversational speech has significant variation in prosody, and can be built by adapting read speech systems with a some amount of prosodically rich data \cite{bhagyashree_asru_2021, ishika_2023}. Hence, we collect a limited amount of expressive speech data.
\end{itemize}



The rest of the paper is organised as follows. Section \ref{sec:related} describes the major TTS datasets available for Indian languages. The proposed unified framework for data collection is presented in Section \ref{sec:proposed}. Related experiments and results are discussed in Section \ref{sec:experiments}. The work is concluded in Section \ref{sec:conclusion}.

\section{Related Work on existing Indic TTS datasets}
\label{sec:related}

We discuss a few of the freely available large-scale TTS datasets for Indian languages:
\begin{itemize}
    \item IndicTTS database: TTS datasets for 13 Indian languages and corresponding Indian English nativities were collected as part of a consortium project funded by the Ministry of Electronics and Information Technology, Government of India \cite{ArunResources2016}. 
    Around 40 hours of data is provided for each language, with 20 hours of native and 20 hours of Indian English audio data recorded by a male and a female speaker. This database is popular with speech researchers working with Indian languages \cite{Baljekar2018, ai4bharat_TTS_2023}. Despite collecting studio-recorded data, the datasets have a few issues. These issues include data collection across different studio environments, recording variability across sessions and incomplete sentences. The performance of systems trained with these datasets varies across languages/datasets, as evidenced by the subjective evaluations in \cite{Prakash2023ASRU}. This motivates the need for a uniform approach to collecting TTS datasets.
    \item SYSPIN \footnote{\url{https://syspin.iisc.ac.in/}} is a large-scale initiative undertaken to collect 40 hours of TTS data in each of the 9 languages-- Bhojpuri, Bengali, Chhattisgarhi, Hindi, Kannada, Maghadi, Maithili, Marathi and Telugu. As part of the LIMMITS challenges 2023 \cite{LIMMITS_2023} and 2024 \footnote{\url{https://sites.google.com/view/limmits24/}}, TTS data has been released for six of these languages (Bengali, Chhattisgarhi, Hindi, Kannada, Marathi and Telugu) and Indian English. However, as reported by the challenge organisers, these datasets consist of many errors including distortion in the audio, wrong pronunciation of words, mismatch between text and audio and transcription in some other language\footnote{\url{https://sites.google.com/view/syspinttschallenge2023/dataset/tts-training-data?authuser=0}}.
    \item Google has released around 2.5 hours of verified high-quality multi-speaker data in 6 Indian languages, collected by crowd-sourcing \cite{hemonry2021opensource}. However, the audio is not recorded by professional speakers and is prone to variability across speakers. 
    \item A parallel and ongoing initiative to collect TTS datasets for Indian languages, Rasa, was published recently \cite{rasa24_interspeech}. The rasa dataset consists of studio-recorded audio by professional speakers of 3 Indian languages (Assamese, Bengali, Tamil), comprising around 10 hours of read speech (neutral data) and 1-3 hours each for 6 emotions. The aim of the work is to build expressive TTS systems. This resource has now been extended to cover 6 more Indian languages (minimum of 20 hours per dataset)\footnote{\url{https://rasa.ai4bharat.org/}}. The data collection protocol is mostly similar to that presented in \cite{ArunResources2016} and mainly differs in the text collection aspect. In our work, we also factor in issues of weak phone coverage and per-word syllable limit during data collection.
\end{itemize}

In the current work, we attempt to provide a high-quality foundation dataset for TTS training, on top of which other datasets \cite{rasa24_interspeech, indicvoices2024} can be used for fine-tuning or augmented based on the required application. In this context, the focus of our work is on proposing a unified protocol for TTS data collection. 




\section{Proposed approach for data collection}
\label{sec:proposed}

The proposed unified protocol for collecting the IndicTTS 23 database is presented here.

\subsection{Studio-recording}
The data for all the 22 languages + Indian English is recorded in a studio. The speech is recorded at a sampling rate of 48 kHz with 16-bit precision. The specifications are maintained throughout all the languages. The recording is carried out without any background noise, reverberations, reflections and echoes.  
\subsection{Voice talent selection and recording}
For each language, a male and a female voice is selected from 3-4 professional Voice Talents (VTs). The best one is chosen based on the voice quality, pronunciation, and amenability to stretching/compression. Voice talents are instructed to familiarise themselves with the text and to carefully read them while maintaining a constant speed and syllable rate. The voice talent must look forward; the head and chin positions should not change while recording. The voice talents are provided with a 15 minute break for every 45 minutes of recording. The audio is checked after recording for correctness and re-recorded if necessary. The recording of the VT is also checked before every session to maintain uniformity in recordings across sessions.

Each speaker per language records 10 hours each of native language and Indian English (total of 40 hours for each of the 22 languages). This poses a challenge, especially for low-resource languages, as it is difficult to find a voice talent fluent in both the native language and English. The 10-hour data is split into two categories-- 9 hours of read speech (neutral prosody) and 1 hour of children's story (expressive speech). 

The data collection process is in accordance with the terms approved by the Ethics Committee of IIT Madras. A consent form is obtained from the voice talents by confirming their acceptance to not only record the voice but also release it under CC BY 4.0 license.


\subsection{Text selection and curation}
The text data is sourced from the web, and LDCIL corpus\footnote{\url{https://data.ldcil.org/text-raw-corpus}}, which encompasses data from diverse domains such as literature, science, newspapers, websites and books. The text is collected with the following criteria:

\begin{itemize}
\item \textit{Sentence length}: The sentences with typically 5-15 words are chosen from the raw corpus to preserve the rate of speech delivery.  Voice talents (however well-trained) have a tendency to read fast when the sentence is long. The tendency among speakers is to start with a lower syllable rate and end with a higher syllable rate.   
\item \textit{Syllable coverage}: The text is chosen in such a way that it contains many  unique syllables.  Syllable distribution in Indian languages have a Zipf  distribution \cite{AnushaIS16}.  Nonetheless, the tails are very long.
\item \textit{Weak phone coverage}: The text is chosen such that it contains all the phonemes of the specific language. The distribution of 20 least occurring phones in the text is then plotted, and based on this, the text is augmented such that at least 50 sentences containing these weak phones are present in the text. Figure \ref{fig:weak_phone_dist} shows the distribution of phones in the Nepali text: (a) in the initial set, and (b) after weak phone augmentation.

\begin{figure}[!h]
 \centering
 \includegraphics[width = 0.9\linewidth]{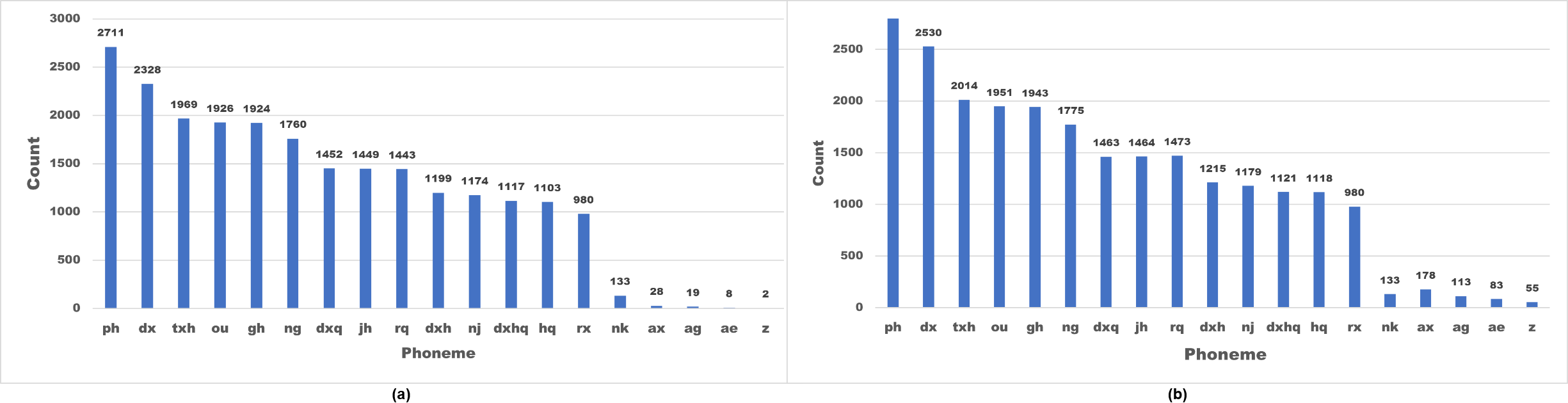}
 \caption{Distribution of phones in Nepali text: (a) initial set (b) post weak phone augmentation}
  \label{fig:weak_phone_dist}
\end{figure} 

\item \textit{Syllable limit per word}: Owing to agglutination, words in Indian languages tend to be long, and can contain even 10 syllables in a word. When such words are read by a voice talent, the syllables tend to be shortened \cite{AnushaIS16} leading to poor quality synthesis.  An effort was made to ensure that text contained at most 4-5 syllables per word.
\end{itemize}

Once the initial text set is collated, it is then carefully curated manually before recording. The correctness of the structure of the sentence is verified, and incomplete sentences and repeated phrases are corrected. Uncommon words and phrases are removed. Text normalization is carried out for sentences with numbers, dates and symbols. Abbreviations,  and acronyms are replaced by their long forms. Punctuation marks, except commas and full stops, are removed from the text. Spelling mistakes and incorrect splitting of words or sentences are also corrected. Words that are long (in terms of number of syllables) are replaced by appropriate words with fewer syllables.  The text is also checked for sensitive content, namely, violence, political, and communal content.

\subsection{Data verification}

The voice quality and pronunciation are checked manually, and the syllable rate at the phrase level is computed. The data verification process is carried out by language experts. Verification of data for 10 hours takes about 20 days at 6-8 hours/day.

The typical value of mean syllable rate  for the audio is 6-8 syllables/second which is considered standard for professional speakers. In Table I, the average syllable rate per utterance is calculated by taking 100 random audio files from Kannada male datasets of IndicTTS23, IndicTTS and Rasa databases, respectively. 

\begin{table}[h!]
\centering
\caption{Average syllable rate per utterance}
\label{tab:syll_rate}
\begin{tabular}{|l|l|l|l|}
\hline
Dataset               & \multicolumn{1}{c|}{IndicTTS23} & \multicolumn{1}{c|}{IndicTTS} & \multicolumn{1}{c|}{Rasa} \\ \hline
Average syllable rate & 7.70$\pm$0.95                       & 7.71$\pm$1.14                     & 8.32$\pm$1.05                 \\ \hline
\end{tabular}
\end{table}


\begin{figure}[!h]
 \centering
 \includegraphics[width = \linewidth]{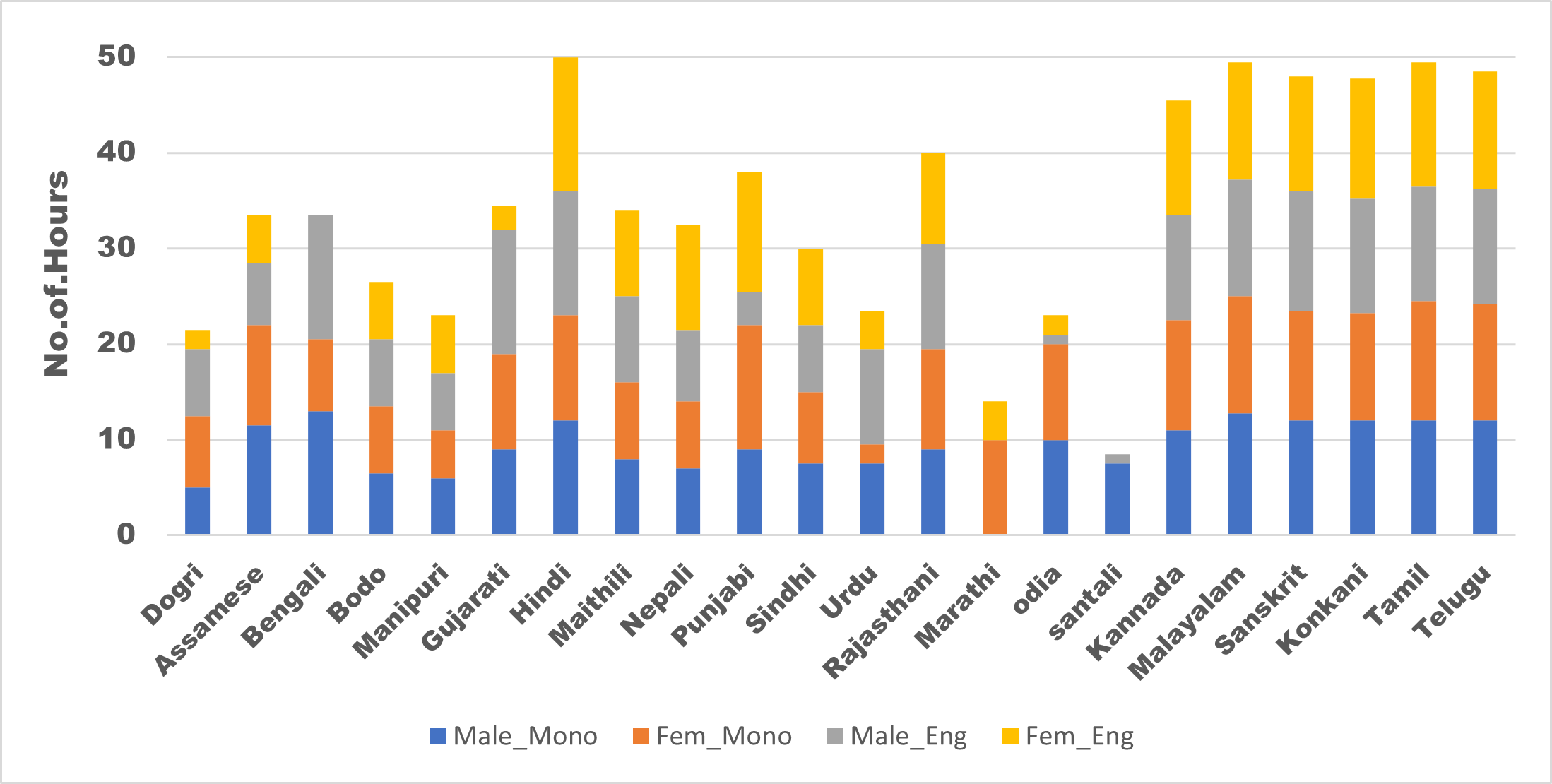}
 \caption{Data collected across languages}
  \label{fig:data_bar}
\end{figure} 



Figure \ref{fig:data_bar} shows the data collected so far, amounting to 765 hours. The data collection process is still ongoing for a few languages \footnote{The IndicTTS23 database will be made available soon at \url{www.iitm.ac.in/donlab/tts/}}. 

\section{Experiments \& results}
\label{sec:experiments}

An attempt is made quantify the quality of datasets collected for a few languages using original recorded audio, and audio generated by different TTS systems and datasets\footnote{Audio samples: \url{https://nltm.iitm.ac.in/indictts23/}}.


\subsection{Comparison of ground-truth recordings}
The original recorded audio collected from IndicTTS23, IndicTTS and Rasa databases were compared for a few languages. A mean opinion score (MOS) like\footnote{We refer to the MOS-like score in the paper as the conventional MOS score compares synthesis quality of systems trained on the same dataset.} test was performed to assess the quality of recorded audio across both datasets. Native listeners were asked to rate the quality of the audio based on the naturalness of speech (1-5 scale, 5 being the best). 8 original recorded audio files from each dataset were presented to the evaluators in random order. Table \ref{tab:mos_gt} presents the MOS-like scores for a few languages. For Hindi and Kannada female, the IndicTTS23 datasets have a higher score compared to the other two datasets. The higher score for the Hindi male dataset from the IndicTTS database is largely due to the superior voice quality of the professional speaker (based on feedback from participants). The quality for Telugu male datasets are rated similarly. 

\begin{table}[h!]
\centering
\caption{MOS-like scores of original recordings across different datasets (NA: not available) No. of evaluators given in ()}
\label{tab:mos_gt}
\begin{tabular}{|l|c|c|c|}
\hline
Dataset            & IndicTTS23 & IndicTTS  & Rasa      \\ \hline
Hindi female (10)  & \textbf{4.58$\pm$0.57}  & 4.22$\pm$0.83 & NA         \\ \hline
Hindi male (10)    & 4.40$\pm$0.49  & \textbf{4.57$\pm$0.58} & NA         \\ \hline
Kannada female (10) & \textbf{4.16$\pm$0.45}  & 3.42$\pm$1.00 & 3.91$\pm$0.86 \\ \hline
Telugu male (14)    & \textbf{4.66$\pm$0.40}  & 4.64$\pm$0.38  & NA         \\ \hline
\end{tabular}
\end{table}

\subsection{Mel-cepstral distortion (MCD) scores of systems trained with IndicTTS23 datasets}

A FastSpeech2+HiFi-GAN \cite{hifigan_NEURIPS2020} system was trained for 4 datasets from the  IndicTTS23 database. We evaluated the standalone performance\footnote{Ground-truth audio files for parallel text are not available across different databases (IndicTTS23/IndicTTS/Rasa).} of these systems using MCD scores. MCD score is an objective measure that determines the distortion in mel-cepstral features of a synthesised speech compared to a reference audio \cite{MCD}. For the dataset, 20 audio files from the test (not seen during training) were considered for evaluation. MCD scores of the synthesised audio files are given in Table \ref{tab:mcd}. 
\vspace{-0.1cm}
\begin{table}[h!]
\centering
\caption{MCD scores corresponding to TTS systems trained with IndicTTS23 dataset}
\label{tab:mcd}
\begin{tabular}{|l|c|c|c|}
\hline
Dataset       & MCD & Dataset & MCD \\ \hline
Hindi female  &    6.32$\pm$0.32      &    Sanskrit male      &    6.30$\pm$0.76      \\ \hline
Nepali female   &    5.53$\pm$0.41      &   Telugu male       &    5.35$\pm$0.55      \\ \hline
\end{tabular}
\end{table}

\subsection{Comparison of TTS systems trained using different datasets}

Audio synthesised from 3 different systems trained on different datasets are compared. The three systems are:
\begin{itemize}
    \item System A: FastSpeech2+HiFi-GAN system trained on the IndicTTS23 datasets.
    \item System B: FastSpeech2+HiFi-GAN system trained on the IndicTTS datasets.
    \item System C: FastPitch \cite{fastpitch}+ HiFi-GAN system trained on the IndicTTS datasets in \cite{ai4bharat_TTS_2023} \footnote{\url{https://ai4bharat.iitm.ac.in/areas/model/TTS/IndicTTS}}.
\end{itemize}

Evaluations are conducted with 4 datasets-- Hindi female, Telugu male, Nepali female and Sanskrit male. Results of MOS-like tests for Hindi and Telugu are presented in Table \ref{tab:mos_synth}. The performance of systems is in the following order: System A $>$ System B $>$ System C.

\begin{table}[h!]
\centering
\caption{Comparison of MOS-like scores of audio synthesised by different systems}
\label{tab:mos_synth}
\begin{tabular}{|l|c|c|c|}
\hline
Dataset          & System A & System B  & System C      \\ \hline
Hindi female (10) & \textbf{4.42$\pm$0.47}  & 4.18$\pm$0.59 & 3.23$\pm$0.73 \\ \hline
Telugu male (10)  & \textbf{4.35$\pm$0.65}  & 4.22$\pm$0.87 & 4.09$\pm$0.83 \\ \hline
\end{tabular}
\end{table}

The languages Sanskrit and Nepali are not available in the IndicTTS database, resulting in only System A for these datasets. Instead of conducting a MOS-like test for these languages, a degradation MOS (DMOS) test was conducted. In the DMOS test, the MOS scores are normalised with respect to the ground-truth audio, presented to the evaluators in random order. From Table \ref{tab:dmos_synth}, we see high DMOS scores for System A, indicating human-like audio quality. The DMOS score of Sanskrit is greater than 5, as more than $50\%$ of the evaluators had rated synthesised utterances more than the actual recordings. The feedback given by evaluators was that it was very difficult to distinguish between natural and synthesised utterances. The results of Nepali are indicative and not conclusive due to the lack of evaluators. The evaluations conducted on ground-truth and synthesised audio highlight the necessity of collecting high-quality TTS datasets for training.

\begin{table}[h!]
\centering
\caption{DMOS scores of System A ( IndicTTS23 datasets)}
\label{tab:dmos_synth}
\begin{tabular}{|l|c|}
\hline
Dataset            & System A  \\ \hline
Sanskrit male (11) & 5.01$\pm$0.36 \\ \hline
Nepali female (2)  & 4.89 \\ \hline
\end{tabular}
\end{table}

\section{Conclusion}
\label{sec:conclusion}

In this paper, we have presented a uniform approach to collecting TTS datasets for 22 Indian languages + Indian English. We have highlighted the importance of collecting studio-recorded and accurately transcribed data for training, emphasising the need for carefully looking at aspects of speaker selection, text collection and data verification. This data collection protocol has also been extended to collecting expressive speech. The datasets collected will be made available as part of the project. 



\section*{Acknowledgment}

This work has been undertaken as part of the project ``Speech Technologies in Indian Languages'' (SP/21-22/1960/CSMEIT/003119), funded by MeitY, GoI. The authors would like to thank various people for helping with data verification and subjective evaluations.

\bibliographystyle{IEEEbib}
\bibliography{references}

\begin{thebibliography}{10}

\bibitem{tacotron2}
Jonathan Shen, Ruoming Pang, Ron~J. Weiss, Mike Schuster, Navdeep Jaitly, Zongheng Yang, Zhifeng Chen, Yu~Zhang, Yuxuan Wang, Rj~Skerrv-Ryan, Rif~A. Saurous, Yannis Agiomvrgiannakis, and Yonghui Wu,
\newblock ``{Natural TTS Synthesis by Conditioning WaveNet on Mel Spectrogram Predictions},''
\newblock in {\em International Conference on Acoustics, Speech and Signal Processing (ICASSP)}, 2018, pp. 4779--4783.

\bibitem{fastspeech2}
Yi~Ren, Chenxu Hu, Xu~Tan, Tao Qin, Sheng Zhao, Zhou Zhao, and Tie{-}Yan Liu,
\newblock ``{FastSpeech 2: Fast and High-Quality End-to-End Text to Speech},''
\newblock in {\em International Conference on Learning Representations, (ICLR)}, 2021, pp. 1--15.

\bibitem{Jeena_transcription_2019}
Jeena~J. Prakash, Golda~Brunet Rajan, and Hema~A. Murthy,
\newblock ``{Importance of Signal Processing Cues in Transcription Correction for Low-Resource Indian Languages},''
\newblock {\em ACM Transactions on Asian and Low-Resource Language Information Processing}, vol. 19, no. 1, pp. 14:1--14:26, 2019.

\bibitem{Thomas1986}
J.~P. Thomas,
\newblock ``{Spatial Vision Then and Now},''
\newblock {\em Vision Research}, vol. 26, pp. 1523--1530, 1986.

\bibitem{censusIndia2011}
{Government of India},
\newblock ``{Census of India 2011},'' \url{https://web.archive.org/web/20180702151828/https://censusindia.gov.in/2011Census/Language_MTs.html}, 2011.

\bibitem{srivastava-etal-2020-indicspeech}
Nimisha Srivastava, Rudrabha Mukhopadhyay, Prajwal K~R, and C~V Jawahar,
\newblock ``{I}ndic{S}peech: Text-to-speech corpus for {I}ndian languages,''
\newblock in {\em Proceedings of The 12th Language Resources and Evaluation Conference}. May 2020, pp. 6417--6422, European Language Resources Association.

\bibitem{ArunResources2016}
Arun Baby, Anju~Leela Thomas, N~L Nishanthi, and Hema~A Murthy,
\newblock ``Resources for {I}ndian languages,''
\newblock in {\em Community-based Building of Language Resources (International Conference on Text, Speech and Dialogue)}, 2016, pp. 37--43.

\bibitem{LIMMITS_2023}
Abhayjeet Singh, Amala Nagireddi, Deekshitha G, Jesuraja Bandekar, Roopa R, Sandhya Badiger, Sathvik Udupa, Prasanta~Kumar Ghosh, Hema~A Murthy, Heiga Zen, Pranaw Kumar, Kamal Kant, Amol Bole, Bira~Chandra Singh, Keiichi Tokuda, Mark Hasegawa-Johnson, and Philipp Olbrich,
\newblock ``{Lightweight, Multi-Speaker, Multi-Lingual Indic Text-to-Speech},''
\newblock in {\em IEEE International Conference on Acoustics, Speech and Signal Processing (ICASSP)}, 2023, pp. 1--2.

\bibitem{rasa24_interspeech}
Praveen {Srinivasa Varadhan}, Ashwin Sankar, Giri Raju, and Mitesh~M Khapra,
\newblock ``{Rasa: Building Expressive Speech Synthesis Systems for Indian Languages in Low-resource Settings},''
\newblock in {\em Interspeech}, 2024, pp. 1830--1834.

\bibitem{bhagyashree_asru_2021}
Bhagyashree Mukherjeee, Anusha Prakash, and Hema~A. Murthy,
\newblock ``{Analysis of Conversational Speech with Application to Voice Adaptation},''
\newblock in {\em IEEE Automatic Speech Recognition and Understanding Workshop (ASRU)}, 2021, pp. 765--772.

\bibitem{ishika_2023}
Ishika Gupta and Hema~A Murthy,
\newblock ``{E-TTS: Expressive Text-to-Speech Synthesis for Hindi using Data Augmentation},''
\newblock in {\em International Conference on Speech and Computer (SPECOM)}, 2023, pp. 243--257.

\bibitem{Baljekar2018}
Pallavi Baljekar, Sai~Krishna Rallabandi, and Alan~W Black,
\newblock ``{An Investigation of Convolution Attention Based Models for Multilingual Speech Synthesis of Indian Languages},''
\newblock in {\em INTERSPEECH}, 2018, pp. 2474--2478.

\bibitem{ai4bharat_TTS_2023}
Gokul~Karthik Kumar, Praveen S~V, Pratyush Kumar, Mitesh~M. Khapra, and Karthik Nandakumar,
\newblock ``Towards building text-to-speech systems for the next billion users,''
\newblock in {\em IEEE International Conference on Acoustics, Speech and Signal Processing (ICASSP)}, 2023, pp. 1--5.

\bibitem{Prakash2023ASRU}
Anusha Prakash, Srinivasan Umesh, and Hema~A. Murthy,
\newblock ``{Towards Developing State-of-The-Art TTS Synthesisers for 13 Indian Languages with Signal Processing Aided Alignments},''
\newblock in {\em IEEE Automatic Speech Recognition and Understanding Workshop (ASRU)}, 2023, pp. 1--8.

\bibitem{hemonry2021opensource}
Fei He, Shan-Hui~Cathy Chu, Oddur Kjartansson, Clara Rivera, Anna Katanova, Alexander Gutkin, Isin Demirsahin, Cibu Johny, Martin Jansche, Supheakmungkol Sarin, and Knot Pipatsrisawat,
\newblock ``{Open-source Multi-speaker Speech Corpora for Building Gujarati, Kannada, Malayalam, Marathi, Tamil, and Telugu Speech Synthesis Systems},''
\newblock in {\em Proceedings of the 12th Conference on Language Resources and Evaluation (LREC 2021)}. European Language Resources Association (ELRA), 2021, pp. 6494--6503.

\bibitem{indicvoices2024}
Tahir Javed, Janki Nawale, Eldho George, Sakshi Joshi, Kaushal Bhogale, Deovrat Mehendale, Ishvinder Sethi, Aparna Ananthanarayanan, Hafsah Faquih, Pratiti Palit, Sneha Ravishankar, Saranya Sukumaran, Tripura Panchagnula, Sunjay Murali, Kunal Gandhi, Ambujavalli R, Manickam M, C~Vaijayanthi, Krishnan Karunganni, Pratyush Kumar, and Mitesh Khapra,
\newblock ``{IndicVoices: Towards Building an Inclusive Multilingual Speech Dataset for Indian Languages},''
\newblock in {\em Findings of the Association for Computational Linguistics ACL 2024}, 2024, pp. 10740--10782.

\bibitem{AnushaIS16}
Anusha Prakash, Jeena~J. Prakash, and Hema~A. Murthy,
\newblock ``{Acoustic Analysis of Syllables Across Indian Languages},''
\newblock in {\em INTERSPEECH}, 2016, pp. 327--331.

\bibitem{hifigan_NEURIPS2020}
Jungil Kong, Jaehyeon Kim, and Jaekyoung Bae,
\newblock ``{HiFi-GAN: Generative Adversarial Networks for Efficient and High Fidelity Speech Synthesis},''
\newblock in {\em Advances in Neural Information Processing Systems}, 2020, vol.~33, pp. 17022--17033.

\bibitem{MCD}
R.~{Kubichek},
\newblock ``Mel-cepstral distance measure for objective speech quality assessment,''
\newblock in {\em Proceedings of IEEE Pacific Rim Conference on Communications Computers and Signal Processing}, 1993, pp. 125--128.

\bibitem{fastpitch}
Adrian Łańcucki,
\newblock ``{Fastpitch: Parallel Text-to-Speech with Pitch Prediction},''
\newblock in {\em IEEE International Conference on Acoustics, Speech and Signal Processing (ICASSP)}, 2021, pp. 6588--6592.

\end{thebibliography}

\end{document}